\documentclass[%
 reprint,
 amsmath,amssymb,
 aps,
 prl,
]{revtex4-1}

\usepackage{graphicx}
\usepackage{dcolumn}
\usepackage{bm}
\usepackage{units}
\usepackage{subfigure}

\begin{document}


\title{Nanowire array photovoltaics: radial disorder versus design for optimal efficiency}

\author{Bj\"{o}rn C. P. Sturmberg$^1$}
\email{b.sturmberg@physics.usyd.edu.au}
\author{Kokou B. Dossou$^2$, Lindsay C. Botten$^2$, Ara A. Asatryan$^2$, Christopher G. Poulton$^2$, Ross C. McPhedran$^1$ and C. Martijn de Sterke$^1$}
\affiliation{$^1$ CUDOS and IPOS, School of Physics, University of Sydney, 2006, Australia}
\affiliation{$^2$ CUDOS, School of Mathematical Sciences, University of Technology Sydney, Sydney, 2007, Australia}

\date{\today}

\begin{abstract}
Solar cell designs based on disordered nanostructures tend to have higher efficiencies than structures with uniform absorbers, though the reason is poorly understood.
To resolve this, we use a semi-analytic approach to determine the physical mechanism leading to enhanced efficiency in arrays containing nanowires with a variety of radii.
We use our findings to systematically design arrays that outperform randomly composed structures.
An ultimate efficiency of $23.75\%$ is achieved with an array containing $30\%$ silicon, an increase of almost $10\%$ over a homogeneous film of equal thickness.

\end{abstract}

\maketitle

Strongly absorbing nanostructures are of increasing importance in photovoltaics, as solar cell thicknesses are reduced to minimise costs \cite{Lewis2007, Tsakalakos2008, Nozik2010}.
Popular approaches to achieve light trapping in these cells include the use of plasmonic nanoparticles and diffraction gratings, as well as nanostructuring the absorbing layer \cite{Pala2009, Atwater2010, Zeng2008}.
These subwavelength nanostructures allow for the macroscopic light trapping limit to be exceeded \cite{Yu2010, Kosten2011}, and unlike random texturing are compatible with the scale of thin-films.
Here we focus on silicon nanowire (NW) arrays as a general form of two-dimensional nanostructured active layer \cite{Hu2007, Garnett2010, Kelzenberg2010, Gunawan2010}.
These structures allow for the decoupling of the charge carrier and absorption lengths, and provide a strong enhancement of absorptance compared to homogeneous films of equal thickness \cite{Kayes2005,Garnett2008,Catchpole2011odd}.

For NW arrays to achieve optimal photovoltaic efficiency, it is of crucial importance to carefully design their geometric parameters.
To this end parameter searches across radius and period have been carried out \cite{Lin2009, Li2009}, concluding that arrays of period greater than $500~\text{nm}$ with approximately $60\%$ fill fractions maximise efficiency.
Further studies introduced disorder in NW position, radius and length, finding that variations in any and all parameters enhance absorption \cite{Bao2010, Du2011, Lin2011}.
These studies were however unable to elucidate the physical mechanism which is responsible for the increased absorption, or the role played by disorder, as they were solely based on purely numerical simulations.
Here we resolve this issue by clarifying the mechanism for enhanced absorption for the particular case of radius disorder.
Using a mode-based numerical method in combination with a simple analytic approximation \cite{Sturmberg2011,Dossou2012}, we study arrays containing increasing numbers of sublattices of different radius NWs.
This approach shows that the NWs' absorption resonances can be widely and predictably tuned by varying their radii.
Since the absorption resonance depends on NW radius, random arrays with a variety of NW radii have a broader absorption spectrum than arrays with identical NWs, leading to higher photovoltaic efficiency.
However, yet higher efficiency is achieved in arrays in which the NW radii are carefully chosen so as to match the solar irradiance spectrum.

\begin{figure}
\begin{center}
     \includegraphics[width=0.95\linewidth]{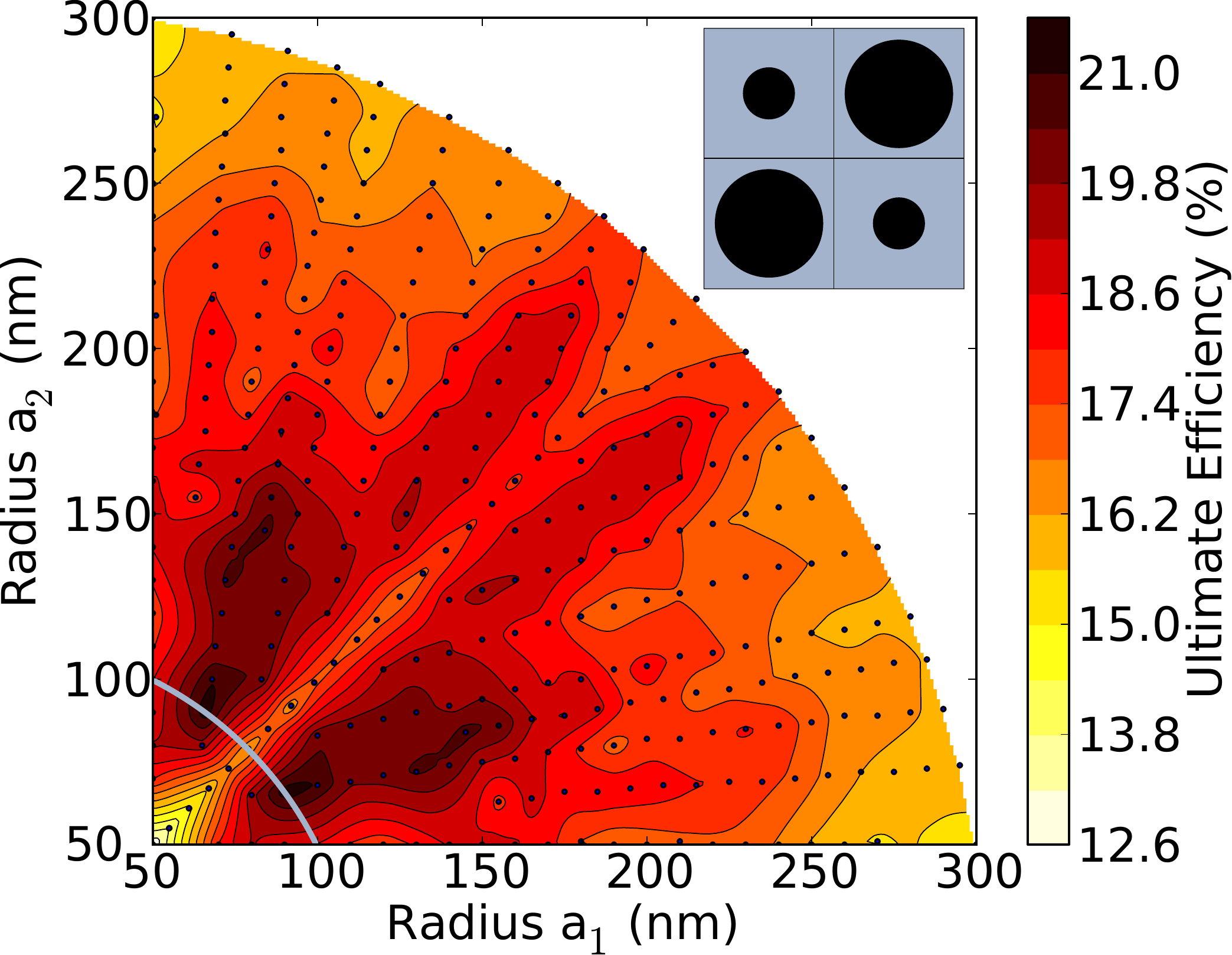}
\end{center}
\vspace{-7mm}
\caption{Ultimate efficiency for $30\%$ silicon NW arrays containing 2 sublattices with differing radius (inset shows a typical unit cell).
Dots indicate simulation results and the arc marks arrays with $d=181~\text{nm}$.}
\label{eta}
\end{figure}

We begin with arrays containing two square sublattices (inset Fig.~1).
Our aim is to compare the efficiency when all NWs have the same radius, to the case in which the NW radii $a_{1,2}$ on the two sublattices differ while keeping the fill fraction $f\equiv\pi(a_1^2+a_2^2)/(2d^2)$ constant at $f=30\%$.
We found this fill fraction to be approximately optimal because it allows for the incorporation of a greater degree of disorder than the $60\%$ found to be optimal for uniform arrays.
Our results are compiled in Fig.~1, which shows the ultimate efficiency\cite{Shockley1961} $\eta$ for NW radii varying between $50 \leq a_1,a_2 \leq 300~{\rm nm}$.
The distance between adjacent NWs $d$ is varied between $115 \le d \le 492~\text{nm}$ so as to keep $f$ constant.
The silicon NW arrays have a thickness (NW height) of $2330~\text{nm}$ and are placed on a SiO$_2$ substrate.
In this figure arrays of equal period lie on circular arcs. Though we see substantial variation of $\eta$ with lattice spacing $d$, the key observation for the present purpose is that for {\sl any} given lattice spacing, the introduction of radial variation, {\sl i.e.}, moving off the diagonal $a_2=a_1$, enhances the efficiency.
This indicates that the effect of introducing radial disorder is universal and does not depend on the lattice spacing.
We therefore henceforth present results for $d=300$~nm only, though we obtained similar results for $d=150$~nm and $600$~nm.
In the case of $d=181~\text{nm}$, which is marked in Fig.~1, $\eta$ increases from $16.2\%$ ($a_1=a_2=79~\text{nm}$), to $21.4\%$ ($a_1=90~\text{nm}, a_2=66~\text{nm}$).

\begin{figure}
\begin{center}
     \includegraphics[width=1.0\linewidth]{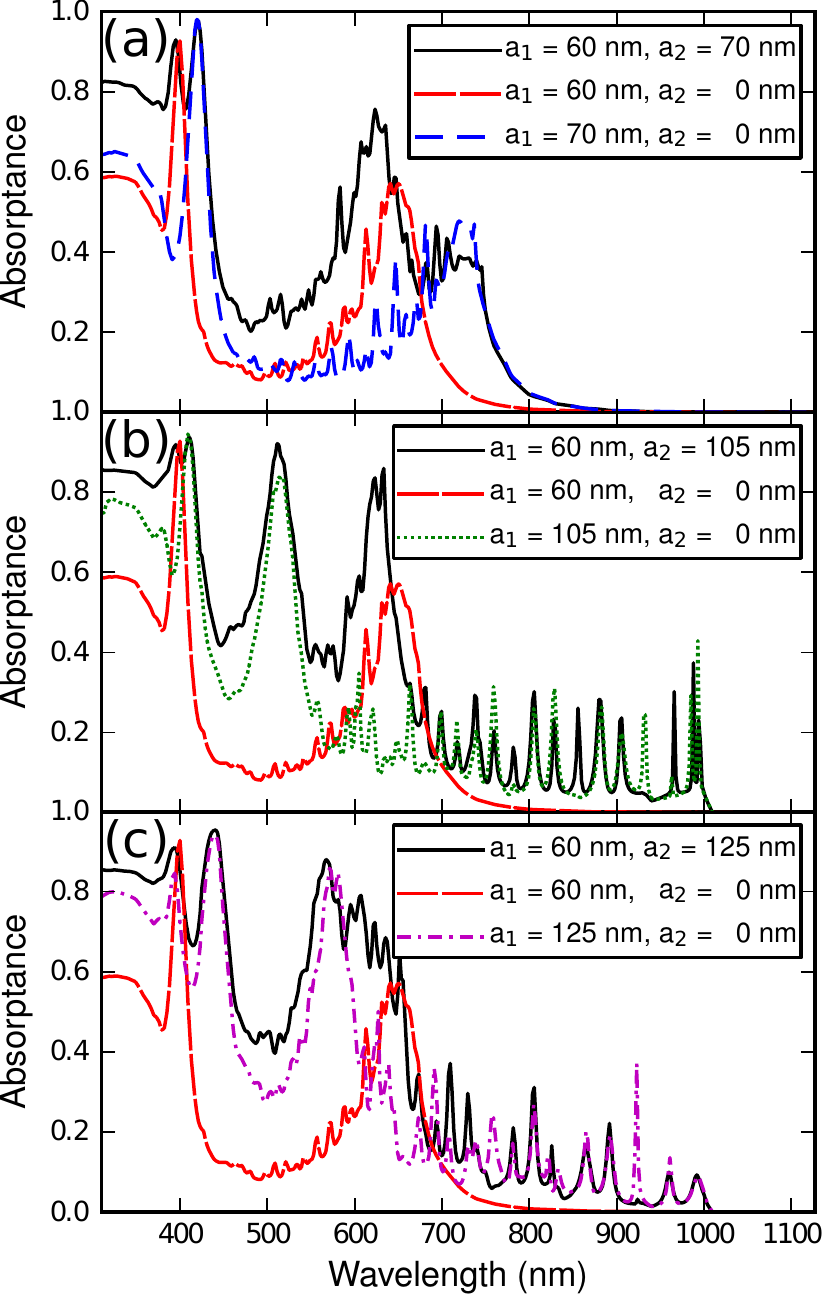}
\end{center}
\vspace{-7mm}
\caption{Absorption spectra of complex arrays containing two radii, where $d=300~\text{nm}$, $a_1=60~\text{nm}$ and $a_2=70, 105, 125~\text{nm}$ in (a), (b) and (c) respectively.
Broken curves show the spectra of individual sublattices.}
\label{overlay2}
\end{figure}

To analyse this enhancement we plot the absorption spectra of complex arrays with $d=300~\text{nm}$ composed of two sublattices: $a_1=60~\text{nm}$ and $a_2=70, 105, 125~\text{nm}$ in Figs.~2 (a)-(c) respectively (black curves).
Also plotted are the absorption spectra of each sublattice separately, {\sl i.e.}, where NWs of one radius have been removed such that $a_1=60, 70, 105, 125~\text{nm}$, $a_2=0~\text{nm}$ (coloured broken curves).
We observe that the superposition of the individual sublattice's spectra is a good approximation to the complex array's spectrum.
This is also the case for the dispersion diagrams (not shown), where the modes of the array containing both radii are the combination of each sublattice's modes in only slightly altered form.
Examining the modal field plots at the absorption peaks confirms that at these wavelengths the fields of the complex array are predominantly concentrated within the NWs of the respective sublattice.

Before examining the superposition phenomenon in detail we summarise the results of our modal analysis of regular periodic NW arrays \cite{Sturmberg2011}.
Our central finding was that NW array absorption spectra are determined predominantly by a small set of \textit{Key Modes} (KMs).
These are ideally suited to strong absorption due to their confinement of light within the absorbing media, low group velocity, and efficient coupling to the incident fields.
They also exhibit Fabry-P\'{e}rot resonances.
The strong dispersion of NW arrays however means that KMs are highly absorbing over only a limited wavelength range of tens of nanometers (see Fig.~2).
The long-wavelength edge of the absorption peak is given by the KM's cut-off wavelength, where its complex valued propagation constant becomes predominantly (due to weak material loss) real and positive.
Using a dipole approximation for the KMs, we found a simple expression for their cut-off wavelength \cite{Sturmberg2011}.

Here we extend the dipole model to describe more complex arrays containing multiple sublattices.
This analysis indicates that the cut-offs of the modes of a particular sublattice depend only weakly on the presence of other sublattices.
In turn, this implies that the absorption of each of the sublattices is essentially independent of the presence of the other sublattices, consistent with Fig.~2.
Such weak sensitivity of modes near cut-off to variations in the surroundings has previously been found in studies of microstructured optical fibres \cite{White2002, Litchinitser2003}.

\begin{figure}
\begin{center}
     \includegraphics[width=1.0\linewidth]{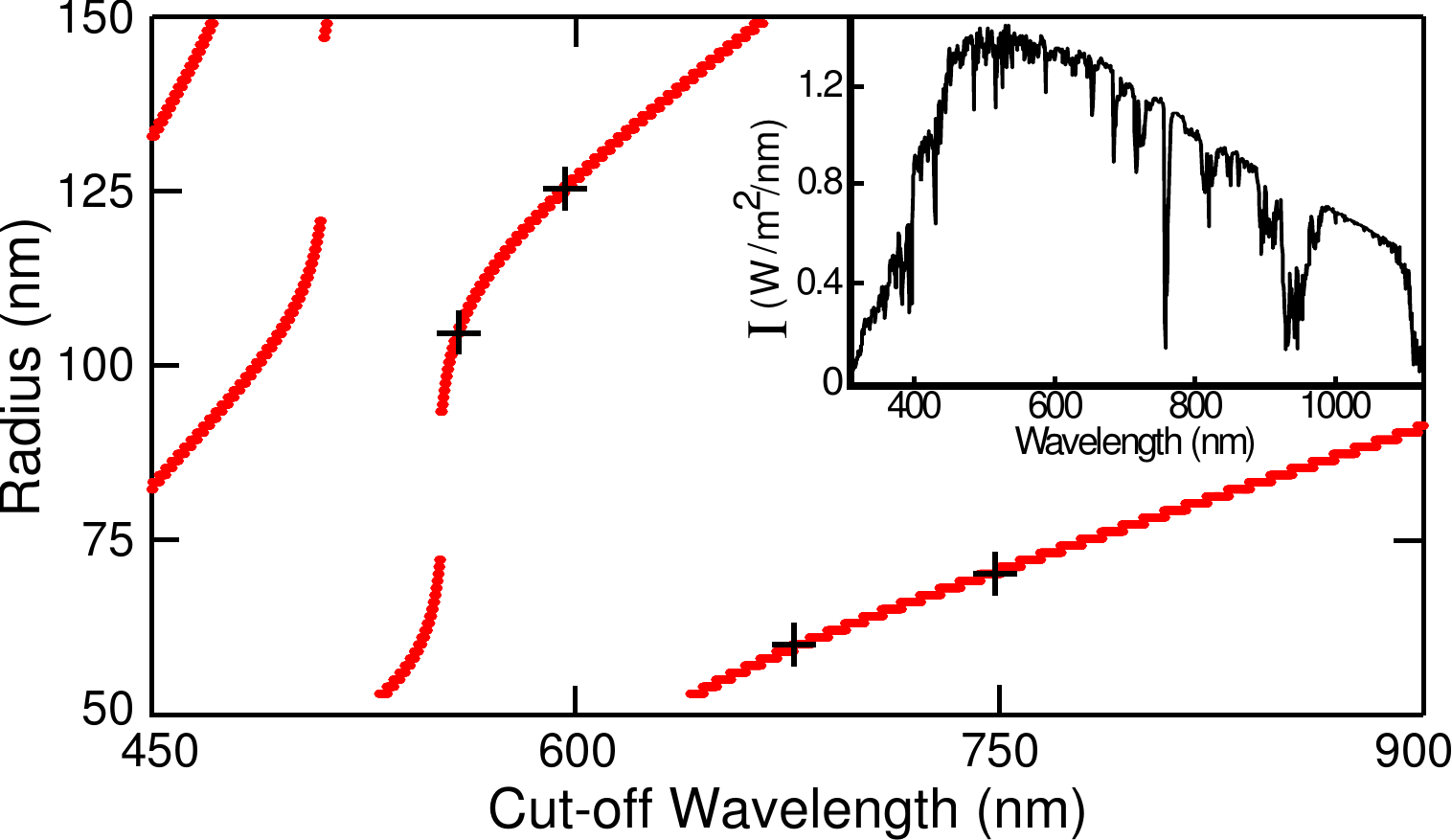}
\end{center}
\vspace{-7mm}
\caption{Key mode cut-off wavelengths for $d=300$~nm as a function of NW radius as calculated by the dipole model. Radii selected are marked by crosses and the inset shows the AM 1.5 solar irradiance spectrum.}
\label{cut}
\end{figure}

Building on this understanding of arrays with two sublattices, we now optimize the absorption of a NW array composed of four sublattices.
To do this we use the dipole model to calculate the KM cut-off wavelengths of a range of NW radii for a set NW spacing.
The result (for $d=300~\text{nm}$) is shown in Fig.~3, where discontinuities occur when no real solutions exist for the cut-off wavelength.
Such a plot is produced in under $10$~seconds, making the dipole model an extremely efficient design tool.

We maximise the ultimate efficiency $\eta$ by selecting KM cut-offs at roughly $70~\text{nm}$ spacing, corresponding to the characteristic width of an absorption peak, across the $450$--$750~\text{nm}$ peak in solar irradiance (inset Fig.~3).
As a proof-of-concept design we select $a=60, 70, 105, 125~\text{nm}$, such that $f=30\%$.
These radii are marked by crosses in Fig.~3.
The absorption spectrum of the designed array (D4), is shown in Fig.~4 (black curve) along with the spectra when all but one of the sublattices are removed (coloured broken curves).
We see that the design process leads to high absorptance across the target wavelength range and that the absorptance is well approximated by the superposition of the 4 sublattice spectra.
Consistent with our earlier argument, permutation of the positions of the 4 NWs does not alter the absorption spectra discernibly.
The ultimate efficiency achieved with D4 is $\eta=22.7\%$, which is a significant increase over a regular array with uniform NWs at equal spacing and fill fraction, for which $\eta=17.6\%$.

\begin{figure}
\begin{center}
     \includegraphics[width=1.0\linewidth]{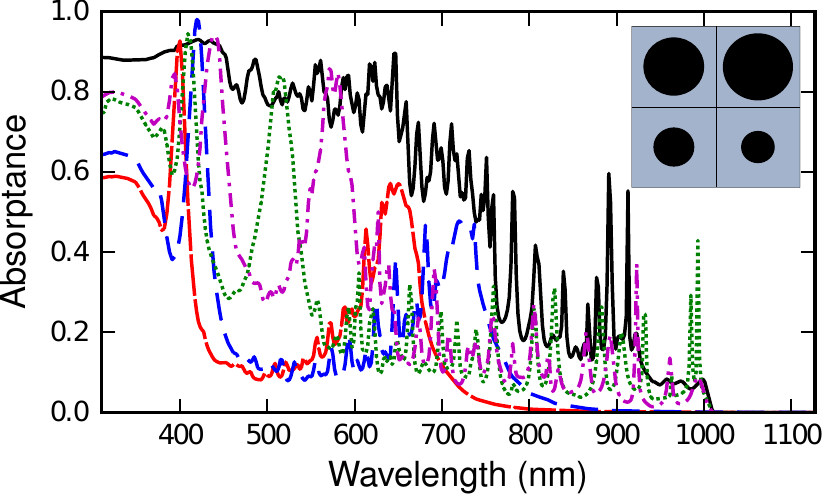}
\end{center}
\vspace{-7mm}
\caption{Absorption spectra of the proof-of-concept designed array D4 (black curve) with spectra for each individual sublattice (coloured broken curves).}
\label{4d}
\end{figure}

To elucidate the effects of disorder we simulate arrays with large supercells in which the radii are chosen randomly.
Specifically, we simulate an ensemble of $100$ arrays with supercells that contain $16$ NWs.
The spacing between adjacent NWs is again $d=300~\text{nm}$ and the radii are selected from a uniform distribution ranging between $50-150~\text{nm}$ (comparable results were obtained for other distributions).
To make a fair comparison we select arrays for which $29.999\%\le f\le 30.001\%$.
Figure~5 shows that the efficiency $\eta$ of the ensemble members (blue bars) varies between $21.4\%$ and $23.6\%$, with an average of $\eta=22.6\%$ (dashed line).
This is a very significant increase over the comparable regular array ($\eta=17.6\%$), but is slightly less than the designed array D4 for which $\eta=22.7\%$ (dot-dashed line).
However, the ensemble members with the highest efficiencies easily outperform the D4 array.
These results indicate that high absorption efficiency requires NWs of different radii, each of which absorbs strongly over a relatively narrow wavelength range.
This suggests that even higher efficiencies can be achieved in carefully designed arrays with sufficient freedom to match the solar emission spectrum more completely.

To this end we design an array containing 16 different radii (D16).
As for the 4 radius case, the dipole model is used to select radii for their KM cut-off wavelengths.
Taking the D4 design as a starting point, we achieved an efficiency of $23.75\%$ with radii of 60, 64, 66, 70, 72, 76, 78, 89, 91, 93, 95, 100, 116, 120, 125, 129~nm.
This efficiency is more than $2.6$ standard deviations above the random arrays' mean, as shown in the histogram on the right of Fig.~5.
Assuming Gaussian statistics, the likelihood of achieving an equal or higher efficiency with a random array is therefore estimated at less than $0.4\%$.
Indeed in Fig.~5 we see that the D16 array outperforms even the best random structure by about $0.15\%$.

\begin{figure}
\begin{center}
     \includegraphics[width=1.0\linewidth]{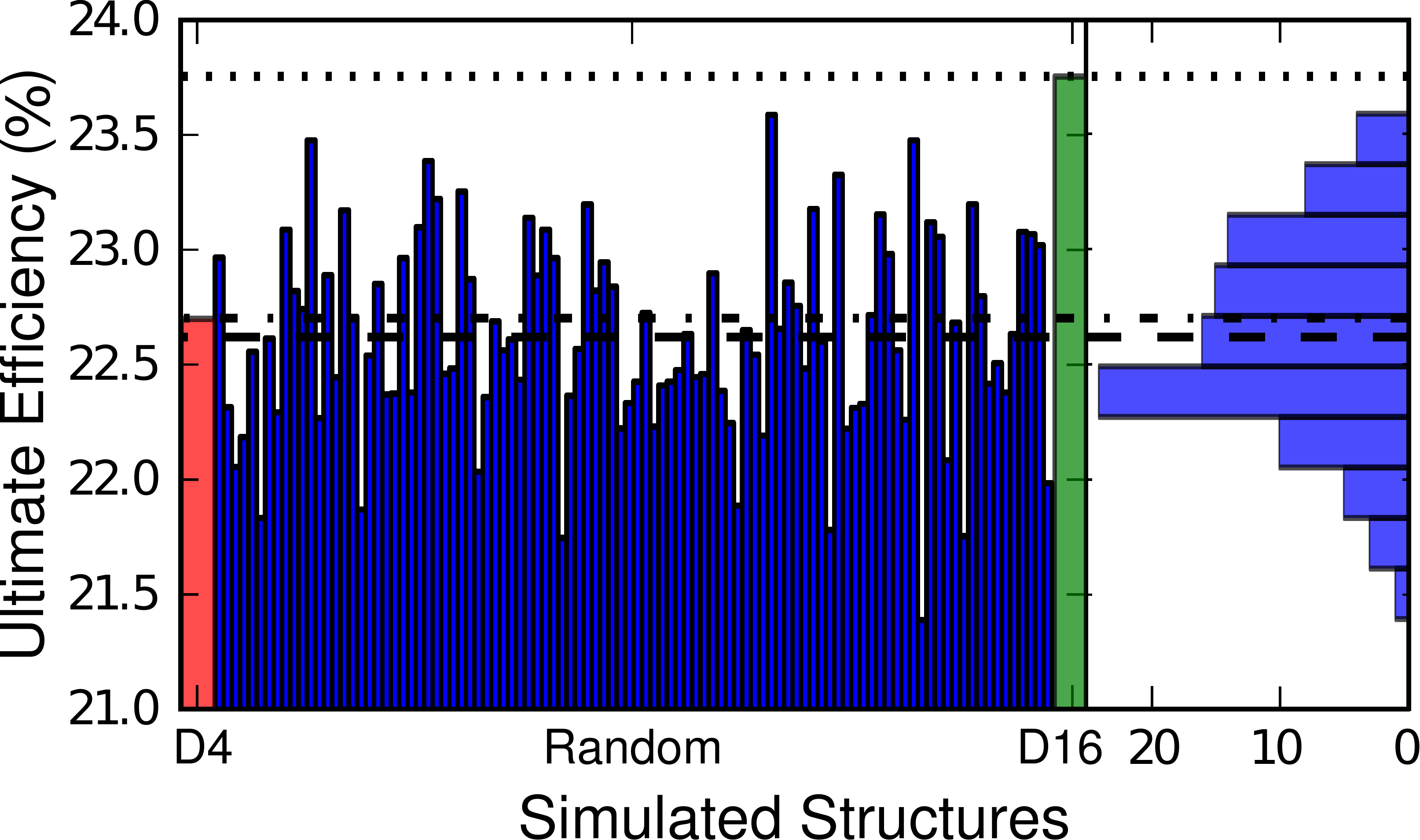}
\end{center}
\vspace{-7mm}
\caption{Ultimate efficiency comparison of $100$ disordered arrays (blue), their average (dashed line), the D4 designed array (red bar, dot-dashed line), and the designed D16 array (green bar, dotted line).
Occurrence histogram of the random structures is shown on right.}
\label{random}
\end{figure}

In summary, we have shown how the inclusion of different radii in NW arrays produces an absorption spectrum that is the superposition of each of the sublattice's spectra.
This dramatically enhances the ultimate efficiency of such structures relative to arrays with NWs of uniform radius.
When the NW radii are selected at random, the average enhancement over uniform arrays is $5\%$.
Furthermore, when the radii are chosen in an informed manner, the efficiency can be increased further.
In total the efficiency of a NW array with $f=30\%$ can thereby exceed that of a homogeneous film by almost $10\%$, rising from $\eta_{\rm HF}=13.8\%$ to $\eta_{\rm D16}=23.75\%$.

\begin{acknowledgments}
We thank Dr Kylie Catchpole and Dr Thomas White for helpful discussions.
This work was supported by the Australian Solar Institute, and the Australian Research Council Discovery Grant and Centre of Excellence Schemes.
Computation resources were provided by the National Computational Infrastructure, Australia.
\end{acknowledgments}

\end{document}